# Reciprocal Collaboration: how lessons from convergence in GLAMs can enhance interdisciplinary AI research

Amber L. Cushing[1], Suzanne Little[2], Giulia Osti[1]

1. School of Information & Communication Studies, University College Dublin, Ireland
2. School of Computing, Dublin City University, Ireland

**Abstract**

The need for collaboration between diverse fields of research is increasingly recognised as important by research funding agencies. A significant driver of this need is the current revolution in artificial intelligence (AI) and related technologies. There is a growing interest in the potential impact of AI in different fields including the methodologies they use and the resulting advances in new knowledge, new access and enhanced productivity. However, there is also a corresponding increase in concern about the fundamentals of AI technologies and the way in which trans and/or interdisciplinary research is approached. The resulting collaboration too often ends up as a one-way street where the domain partner acts only as an information provider. For example, the contribution of the AHSS partner might be limited to providing insight about ethics and/or the technology partner may only provide a service to build applied AI-based solutions. In response to this problem, we propose a *reciprocal* approach to collaboration where both partners seek to understand, cooperate and identify jointly significant impacts. In this paper we explore this relationship between cultural heritage institutions (GLAMs), Arts, Humanities & Social Sciences (AHSS) research and technology-led AI research, especially the impact of current technological advances in AI. Drawing from the history of convergence in GLAM studies, we propose five key practices to form a framework for greater understanding across this divide.

# 1 Introduction

In the past decade, academic research on the topic of artificial intelligence (AI) has expanded dramatically. We use the terms AI and AI technologies to refer broadly to computational approaches – including machine learning (ML), computer vision (CV), natural language processing (NLP), and both traditional statistical and generative methods. Across nearly all field scholars are discussing how to adopt AI, as well as how AI challenges well-established theories and concepts across academic disciplines. As Universities develop Research Centres to establish their presence in AI research and development academic disciplines are identifying how their knowledge and expertise can be relevant in AI development. AI development can be considered a "wicked problem" (Rittel & Webber, 1973) because of the rapid pace of change, the limited understanding of its societal impact and that it exists as something that must constantly be managed and adapted.

Within these academic discussions, interdisciplinary research is often cited for its ability to address "wicked" problems (Crespo Triveño et al., 2026; Devine et al., 2025; Frodeman et al., 2017; Krueger et al., 2025). Beyond traditional fields of computing and engineering, other fields have identified how they can contribute to AI research and development within interdisciplinary collaborations. Authors from the arts, humanities and social sciences (AHSS) disciplines have discussed how their theories and concepts can provide guidance to the fields that develop AI tools (Davies et al., 2025; Jo & Gebru, 2020; Klein et al., 2026; Kommers et al., 2026; Sherman et al., 2024). These authors argue for the integration of AHSS approaches, theories and concepts into AI, ML and CV research and practice. Yet, these scholars also acknowledge that AHSS-based research currently has little uptake in these fields. This points to a broader challenge: although fostering research collaboration across different disciplines has become a major component of science and research policy, it is not easily achieved in practice (Fishberg & Kropp, 2025). By exploring outputs and funding documents, Fisher and Kropp (2025) identified a gap between the ambition for interdisciplinary research and its realisation, stating that knowledge integration often remains superficial.

From the perspective of AHSS researchers, this means their contributions are often limited to providing knowledge that supports science, technology, engineering, maths and medicine (STEMM) researchers to answer disciplinary based questions in their own fields without necessarily being provided an opportunity to contribute to the overall direction of the project (Vienni-Baptista et al., 2024). In contrast, computing experts can often be asked to build applied tools to suit humanities projects, again without an opportunity to explore questions of interest within their own field (Oberbichler et al., 2022). These findings mirror our anecdotal evidence uncovered during discussion in different academic circles: while interdisciplinary collaboration is widely valued, interdisciplinary AI research collaborations often involve significant friction, and leave many participants dissatisfied with previous experiences. There exists a need for models of *how* to enhance AHSS-STEMM interdisciplinary collaboration in the context of technological change. We believe the history of convergence of digital practices in galleries, libraries, archives and museums (GLAM) offers a useful case for thinking about how these interdisciplinary collaborations can develop in practice.

In the 1980s with the rise of personal computing, leaders in library science, archival studies and museum studies—collectively referred to as "LAMs" (the G for galleries came later when museums and galleries were subdivided)—began researching about, writing about, and experimenting with how these separate institutions could harness newly introduced technology to provide users with digital access to cultural heritage materials (Bastian, 2017). More recently, these institutions have worked together to adopt and/or adapt novel technology that utilises large collections of digital materials, including AI technologies. We use the terms "GLAMs" and "GLAM studies" to refer to this site of cross-sector collaboration, rather than pointing to new fields of practice, and by "convergence" we mean the processes through in which different epistemic cultures and philosophies of practice coexist, even as they work toward shared technological and organisational goals.

This paper is structured in three parts: we begin by identifying the operational and structural barriers hindering a meaningful development of interdisciplinary AI research; second, we draw upon the history of convergence in GLAM studies and unfold lessons in translation and reflexive practice; last, we propose a reciprocal approach to collaboration, offering an actionable framework for integrating diverse disciplinary perspectives into AI research and development.

# 2 Strains of interdisciplinary AI research collaboration

## 2.1 Key examples

One of the most notable pieces identifying how AHSS research can be utilised in ML is from Jo and Gebru (2020). The authors argue that as archival studies research has a long history of discussing document collection and annotation and that the theories from this field can be utilised to develop frameworks and procedures to address problems with fairness, accountability, transparency and ethics in machine learning systems. This piece is of note because it was published relatively early in the discussion of AHSS and its relevance to AI and was authored by two researchers with a background in computer science and history. In parallel with this initial reflection, Gebru et al. (2021) introduced *datasheets for datasets,* a human-readable, descriptive approach to documenting the developmental steps of ML datasets. This template was aimed at fostering transparency and accountability between dataset creators and consumers, facilitating the scoping of societal biases encoded in datasets and consequently supporting researchers and practitioners in making informed decisions about dataset-task pairings. Subsequently, adaptations of datasheets-inspired approaches in ML (Mitchell et al., 2019; Pushkarna et al., 2022) have been used by the digital cultural heritage (DCH) domain, expanding on them to pursue complementary objectives. While retaining a focus on contextual documentation and human-readability, the various proposals in this area (Alkemade et al., 2023; Lee, 2025; Luthra & Eskevich, 2024; Maemura & Byrne, 2024) are strongly oriented towards machine-readability, to maintain interoperability with existing information infrastructures. Yet, this development is not unique to the DCH community, but forms part of a cross-pollination process between AHSS and AI/ML documentation approaches, as both are seeking to reconcile human- and machine-readable requirements.

Beyond documentation and DCH, recent scholarship has further articulated what different AHSS disciplines can offer the development of AI. Klein et al. (2025) developed an eight-point list of provocations from the humanities, arguing that humanities scholars have deep experience beyond a general board concept of "ethics" that are relevant in the process of developing generative AI. These include tasks such as developing and preserving archives, creating theories and arguments, understanding complexity performed through a lens that attends to power and culture. Davies et al. (2025) makes a similar argument in relation to the social sciences and the development of the foundation models as part of AI development: social scientists also understand power, and the necessity to attend to potential social impacts in the process of technology development.

Both papers, along with Sherman et al. (2024) argue for collaboration between AHSS and those who develop AI, while there is a growing call for a qualitative shift as AI becomes integrated into infrastructures. This is a need highlighted in the white paper from the Alan Turing Institute's *Doing AI Differently* project (Hemment & Kommers, 2025). The authors state that, to build AI systems capable of meaningful engagement with human culture and its complexity, frameworks supporting the interpretation of the very artefacts that are co-produced with these systems must be in place. According to them, these frameworks can be developed only by integrating AHSS as core disciplines leading this shift, aimed towards what they term as the creation of 'Interpretive AI,' or "systems designed to handle plurality, ambiguity, and contextual meaning as core capabilities" (Hemment & Kommers, 2025, p. 6). Amongst the core innovations, the authors hint at DeepSeek hiring patterns under the CEO Liang Wenfeng, which have prioritised humanities scholars among other non-technical profiles, as playing a crucial role in data evaluation and qualitative model refinement. Hirsbrunner et al. (2024) track down several of these interdisciplinary encounters while arguing for the need for a critical outlook permeating the technical practices in the context of AI, an argument echoing the call for transdisciplinary collaborations to lead the future of AI research (Krueger et al., 2025).

Little research has explored how to operationalise AHSS research concepts for development of AI systems. One example that does address this is Klumbyte et al.'s (2022) account of a set of design workshops for ML and human computer interaction (HCI) students to introduce humanities based intersectional methodologies and generate design approaches informed by these methodologies. They found that students were open to engaging with these concepts, but that the students also experienced tension between the critical concepts and "disciplinary conventions and pragmatic orientations of computing design that is often results-oriented" (p. 1535). While not specifically about AI, Søndergaard et al. (2023) describe how fabulation can be utilised by HCI researchers in participatory design practice by combining facts and speculations. Based on the work of Hartman (2024) critical fabulation involves using critical theory, archival research and fiction to address silences in the documentary record. Søndergaard et al. note that the application of critical fabulation to HCI affords the practice of reflexivity in design, by allowing the researchers to reflect on their own positionality.

## 2.2 Existing difficulties with interdisciplinary AI research

Although interdisciplinary collaboration is widely recognised as essential for AI research, current collaborations often struggle to translate concepts, values and interpretive frameworks across disciplinary boundaries. Fletcher & Lyall (2020) present a case study of an apparently successful yet "unsatisfactory" collaboration between biomedical researchers and social scientists highlighting the misunderstandings and the sense of speaking different languages. Figure I illustrates some of the unidirectional communications and demands that contribute to this breakdown in understanding.

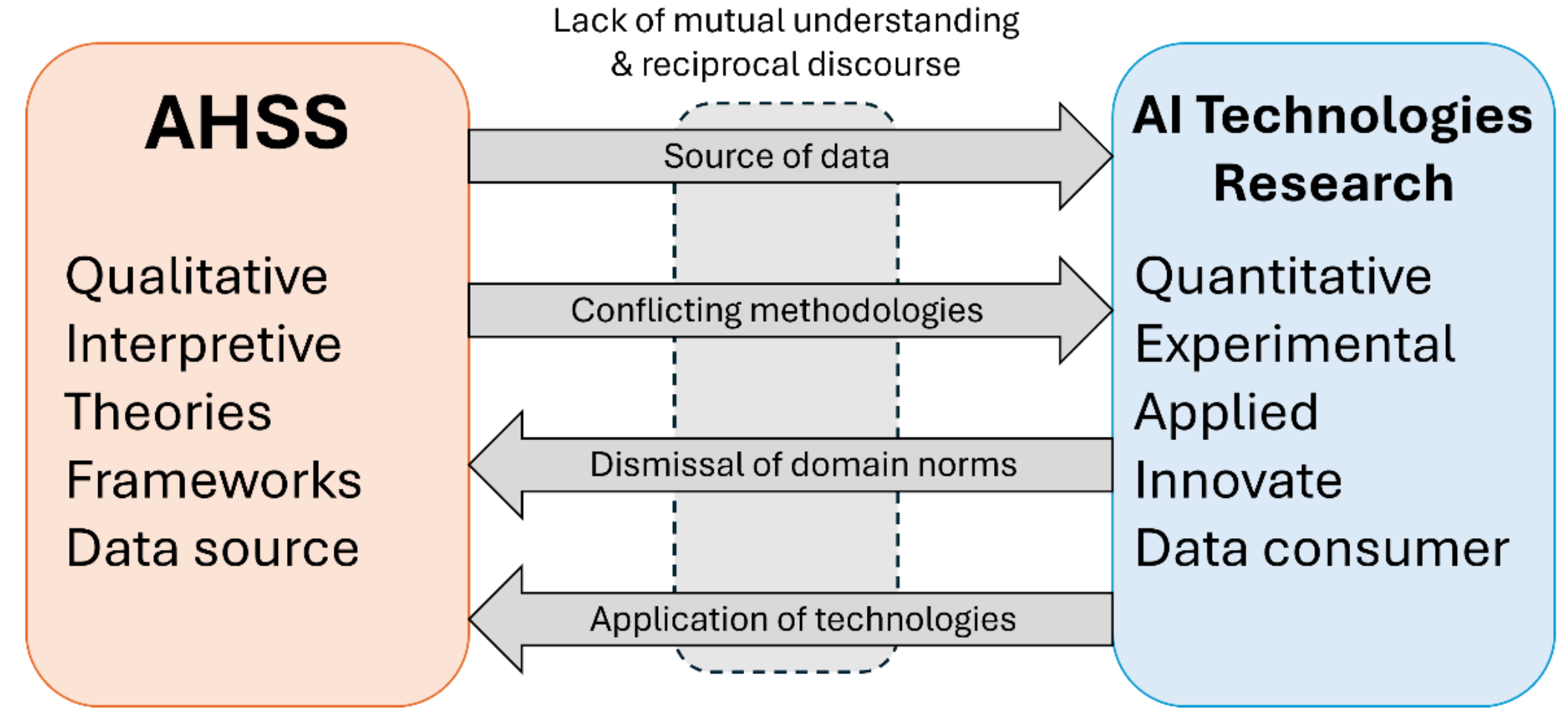

*Figure 1. Common barriers to reciprocal collaboration between AHSS and STEM-led AI researchers*

Academic literature addressing the current dissatisfaction with interdisciplinary AI research collaboration focuses on frustration with what can be referred to as the "parachute approach." This term comes from the environmental science domain, in which academic scientists "parachute" into an area, tell locals about their problems, suggest overly general solutions and then parachute back out (Asase et al., 2022). This approach ignores existing local infrastructure and local expertise on the issue(s) in question and gatekeeps opportunities to provide solutions to those within academia. Existing barriers to interdisciplinary research collaboration can also be described using this parachute metaphor: an academic from outside a specific discipline is asked to "parachute in" to consult on a specific aspect of research and is not invited to provide solutions impacting the overall project.

This is reflected in scholarship about interdisciplinary collaborations. Specifically, Vienni-Baptista et al. (2024) found that AHSS scholars cite several challenges when collaborating with STEMM, including a belief that terms from their own field are "co-opted" without involving AHSS researchers, the frequent ignoring of arts and performance, limited leadership opportunities within research projects, differences in research priorities, conflicting motivations, lack of time to integrate different views within funded projects when outputs are expected to match pre-established timelines, mismatched disciplinary reward structures and incentives, limited institutional support for cross-disciplinary collaboration, and academic tribalism. Klein et al. (2025) echo many of these concerns, citing frustration with lack of opportunities to form equal partnerships within AI research collaborations and a perceived lack of willingness to understand the nuances of different humanities disciplines, beyond philosophy.

While these challenges with interdisciplinary research collaboration are not limited to AI or novel technology, the current focus on AI development and application, alongside standing calls for interdisciplinary collaboration in the funding landscape, has thrown a spotlight on these issues. From one perspective the technology claims to sweep forward, bringing progress, creating efficiencies, and enabling ever greater research achievements and tangible societal and economic impacts. However, in the target domains, the technology can ignore important methodologies, amplify mistrust, bias and unethical behaviour, exploiting data and knowledge without providing true benefits, remaining inextricably rooted in practices that are inherently perceived as negative and problematic (Ciston, 2019). How can genuine interdisciplinary research be conducted where such barriers exist?

# 3 Insights from the history of convergence in GLAM Studies

## 3.1 A brief introduction to convergence in GLAMs

A central mission of GLAMs is to collect, preserve, maintain and make accessible the cultural and intellectual record (Dempsey, 2000; Hughes, 2011). The professionals who execute these tasks do so while valuing equitable access, individual privacy and ethical concern for marginalised voices and the public obligation to maintain history and culture. While their specific values, theories and intellectual constructs may differ, they overlap in their general mission to provide users with access to "universal knowledge" (Bastian, 2017; Marty, 2014).

Practically, convergence is defined as the merging of cultural heritage institutions and/or cultural heritage collections (Bastian, 2017). At a disciplinary level, it involves different cultural heritage institutions working toward a united set of standards and best practices that reflect disciplinary theories and values (Dempsey, 2000; Hughes, 2011). According to Bastian (2017), the convergence of libraries, archives and museums (LAMs) is intertwined with the development of novel technology, originating in 1970s discussions of how information technology could enhance work. Technology was seen as a method to enhance access to the knowledge that LAMs have a mission to preserve and to unite information for the users LAMs continue to have a mission to support (Marty, 2014). Not all LAMs converged and some that converged were disbanded due to changes in governmental policy (Vårheim et al., 2020). An example of GLAM convergence is the merging of different governmental cultural heritage departments into a single unit. This could include moving the professional who work in this unit to the same building or just moving their location within a governmental hierarchy. Within academic research, an example of GLAM convergence is different research labs who seek to conduct research that supports "GLAMs" as opposed to individual galleries, libraries, archives or museums. Within education, training programmes may describe themselves as preparing students to work in GLAMs in contrast to a specific profession.

In the past decade, more research has been published that explored the lasting effects of GLAM convergence. Some findings are overwhelmingly positive, such as high collaboration rates among GLAM professionals: according to Tóth et al.'s (2025) recent survey of GLAM professionals: "Forty-six percent of librarians confirm that they collaborate with archives and/or museums, 60% of archives report collaborating with libraries and museums, and 73% of museologists collaborate with libraries and/or archives" (p. 862). In their study of five memory institutions that converged in New Zealand and Australia, Robinson (2016) found that professionals from different disciplines reported that communications across the different institutions improved in two different ways: formally, through joint meetings and also informally through a growing respect for different types of professionals. Professionals at one institution reported "a culture of mutual respect, teamwork and sense of joint purpose had developed even though all staff members were not co-located in the same office." (p. 147). At another institution, professionals reported "a growing appreciation for other collection areas and the specialist skills of colleagues as a result of the communication necessitated by the converged structure" (p. 147). The professionals who worked in converged institutions were also more likely to report satisfactory collaborations that lead to creative approaches on different exhibits and projects.

However, some research has exposed negative impacts of GLAM convergence: Robinson (2016) notes that convergence does not guarantee improved communication, as administrative structures must be put in place to make professionals aware of their responsibilities and opportunities relating to the convergence whilst simultaneously recognising the value of their specialist expertise. Specifically, if convergence was justified to cull staff roles or as a mere bureaucratic exercise, professionals were likely to be left dissatisfied with the newly converged institution and retreat into their own disciplines, viewing convergence as a threat to their professional identity. Tóth et al. (2025) found that librarians, archivists and curators do not perceive their roles as similar which at first glance may suggest that convergence has not led to a new discipline. The authors are quick to note that the professionals who did not perceive their roles as similar also perceived this fact as a justification to more frequently collaborate because collaborating with professionals in different roles meant that they were more likely to learn new things in the collaboration. These findings suggest that leadership plays an important role in the success of GLAM convergence as leaders clarify structures and standards for communication and that professional education programmes could play a role in training professionals in a new converged discipline. It also suggests that within interdisciplinary collaboration, a fear of losing one's professional identity may not necessarily inhibit willingness to collaborate if common ground can be found–in the case of Tóth et al's survey respondents, the common ground was social mission. While convergence is not without its critics, discussions of the negative and positive impacts of GLAM convergence provides a critical analysis of the reality of the lived experience of different professionals who attempt to practice in converged institutions. These lived experiences provide insight into what works and what does not when different disciplines attempt to converge.

## 3.2 Translating critical work to applied practice via reflexivity

As a discipline, GLAM studies develop concepts and theories and then must translate those theories and concepts for applied practice of access and use of different collections. Often this includes seeking theories and concepts beyond these fields, exploring their relevance and application in GLAMs practice and then investigating how applying these theories and concepts can enhance the practice of collecting, documenting, describing, preserving, accessing and using. For example, Caswell and Cifor's (2016, 2021) application of ethics of care to archival practice has had significant impact on the field of archival studies and serves as an example of how to operationalise feminist ethics in archivist's practices and the process of digitisation. The authors focus on archivists' decision making in the practice of making content digital accessible. In another classic example, Dervin & Dewdney (1986) developed the theory of sensemaking based on communication theory and translated it to the practice of neutral questioning in the reference interview, a common practice performed by a reference librarian. In a recent example, Hicks (2020), a museum curator, applies concepts of decolonisation to aspects of museum practice in *The Brutish Museums: The Benin Bronzes, Colonial Violence and Cultural Restitution*. While the context discussed in these examples of ethics, fairness, representation and access is at the fore, the more latent discussion of the process of translating theory to practice occupies a supporting role. Lengthier discussions of this process are often referenced in methodologies or in research methods training in applied GLAM studies programmes.

These latent discussions of process, considering how critical concepts are translated for applied practice, are needed to enhance application of critical concepts in ML and CV research. GLAM studies research not only explores how to apply critical concepts in practice, it also has a rich history in detailing what content its professionals should learn and why. The work that addresses the education of GLAM professionals, specifically what these professionals should learn and why, complements discussion of applying theory to practice by discussing how to teach new professionals to apply this research in their professional roles. For example, Cushing and Shankar (2019) state that in the specific geographic context of Ireland, most digital preservation experts work individually, versus in a team. As a result, in their educational offering, emphasis was placed on allowing students to consider how to build a virtual community of like-minded practitioners. This and other discussions of teaching provide techniques for translating theory to practice.

One of the methods that guides this application of theory to practice is reflexivity. It involves constant self-reflection and critical self-exploration in all elements of the research process (Corlett & Mavin, 2018). Topically, it involves the individual researcher becoming aware of their position, identity and power within the research process and their relationship to these concepts. According to Finlay (2002), reflexivity can act as a valuable tool to: examine the impact of one's position; promote rich insight by examining interpersonal dynamics; open unconscious motivations and implicit biases in a researcher's approach and evaluate the research process, methods and outcomes. Reflexivity is commonly utilised in GLAM Studies research and practice. For example, the contemporary, librarian, archivist, gallery or museum curator constantly considers their position in relation to the materials they collect, curate, organise and/or preserve and how their position can impact their practice. GLAM professionals also consider how the materials in their care represent and communicate different lived experiences in different contexts and how shared meanings of significance influence their practices. GLAM studies researchers utilise reflexivity in research about how to develop theories, concepts and standards to guide and influence practice.

## 3.3 Technology adoption in the context of convergence

Within the history of GLAM convergence, the adoption of novel technology is contextualised by discussions of convergence of individual institutions into collaborative efforts. In addition to shifts in cultural policy (Tóth et al., 2025), the technology affords this convergence, the approaches to technology adoption cannot be unbraided from parallel discourse about merging values, theories and concepts to form new standards and best practices. According to Rasmussen & Hjørland (2022) providing access to users is not necessarily a technical issue, but it is "intimately connected to an understanding of the value and relevance of what is mediated" (p. 596). In converged GLAMs, technology adoption is contextualised by the acceptance of different approaches to providing access to users driven by a shared value of the collections.

Within that context, there also exists examples of interdisciplinary research projects that detail how to work toward a united goal, flexible enough to meet the needs of all involved. For example, in describing the READ-COOP cooperative model ownership of the artificial intelligence driven Transkribus tool for automatic recognition of handwriting in archival documents, Terras et al. (2025) describe consideration for the collaboration between humanities and computer science scholars. Particularly in the field of digital humanities, research collaborations between STEM authors and humanities authors continues to grow (Ma & Li, 2022). Hansson & Dahlgren (2022) ground their framework for understanding the use of participatory tools in GLAM digitisation projects by pinpointing how values align with different understandings of democracy. The framework guides professionals to consider their approach to the democratic values that support GLAM practice and determine if adoption of certain participatory tools and technological uses align with one's understanding of democratic values. Returning to AI, Cushing & Osti (2023) found evidence of how digital archivists' concern for how novel technology can challenge their concepts of professional expertise, before adopting the technology. In GLAMs, novel technology can prompt professionals to consider not only how to operationalise the technology through tools and workflows, but also to prompt review of core missions, standards and best practices.

In summary, one of the gaps in existing arguments for interdisciplinary AI research is the lack of a roadmap to achieve collaborations that create of new conceptual frameworks, hypothesis, research strategies and approaches to extend beyond them via the process of integration (Vienni-Baptista et al., 2024). GLAM studies research addresses how to translate theories and concepts to applied practice; it plays a significant role in affording the development of these practices and their alignment with jointly developed standards and best practices and demonstrates the significance of well thought out structures that can encourage collaboration in the future and produce creative outcomes.

The history of convergence in GLAMs is reminiscent of the process of integration in interdisciplinary research collaboration. Galleries, libraries, archives and museums identified shared values to create joint standards and best practices to advance the field in the context of technological innovation, all while acknowledging different practices used to maintain the cultural and history record (Bastian, 2017). Convergence was not always a positive experience, but research on the convergence experience provides critical reflection of what can lead to satisfactory outcomes. This critical reflection is often achieved via reflexive practice.

# 4 A reciprocal approach to guide interdisciplinary collaboration in AI research

In the previous section, we identified existing barriers to interdisciplinary AI research collaborations including the often used "parachute approach", lack of integration between researchers from different disciplines and lack of devices to translate AHSS concepts to research on the development and application of AI technologies. We believe that these current barriers to interdisciplinary research can be addressed with the development of a reciprocal approach to interdisciplinary AI research involving the practice of reflexivity to help researchers integrate their disciplinary conventions and communicate more openly with one another about their disciplinary norms. We believe that applying reflexivity to the process of interdisciplinary research collaborations on AI development and application on a variety of different topics can afford improvement on the challenges highlighted by other authors: the low uptake of AHSS concepts and theories in machine learning and computer vision research, frustration with researcher roles in interdisciplinary research collaborations and the need for guidance that translates AHSS research for machine learning and computer vision practice.

**We define a reciprocal approach to interdisciplinary research collaboration as an ongoing co-constituted process in which the identification of written shared values directs decision making within the research collaboration.** Within interdisciplinary research collaboration, reflexivity can be used to uncover the different agendas of each team member in the research collaboration and avoid the biasing of results, findings and outputs toward one discipline and/or perspective (Finlay, 2002).

The reciprocal approach involves constant communication with collaborators about all elements involved in the research process at the highest and lowest levels through a process of mutual reflection (Finlay, 2002). Through this mutual reflection, multiple voices and conflicting positions surface and can be openly considered by the team. It posits that a researcher's experience of collaboration is developed based on interactions with others in a context of research culture, built on shared meanings of different elements involved with research. A reciprocal approach assumes that the research collaboration relationship can shape the findings that are produced in the project.

The reciprocal approach recognises that concerns about deviation from disciplinary norms and threats to concepts of professional identity can represent barriers to satisfactory interdisciplinary collaboration and therefore must be considered and discussed throughout the research collaboration. The reciprocal approach is antithetical to the "parachute" approach described above, in which disciplinary norms associated with novelty, outputs, methods and approaches are ignored and bilateral decision making is limited.

The reciprocal approach affords integration of different disciplinary research questions, approaches, methods and solutions that provide guidance of how to apply and utilise cross disciplinary concepts and theories in different contexts. We believe that this reciprocal approach can be built upon learnings from the history of convergence in GLAMs that addresses the translation of theory to practice and adopting technology in the context of disciplinary and institutional convergence. We suggest using GLAMs as a testbed for interdisciplinary AI research projects because they provide a context with experience in translating concepts for practice to adopt novel technology, and skill and expertise at identifying values to develop joint standards and best practices which can then be translated for endeavours in different scientific contexts. As our focus is on process, we believe that GLAMs provide an environment where the necessary integration process that Vienni-Baptista et al. (2024) describe can be achieved with less friction, allowing us to refine our proposed process.

Below, we provide five key initial points for reciprocal collaboration in practice and ultimately how a reciprocal approach can be applied to enhance interdisciplinary AI research. Grounded in reflexivity as mutual collaboration (Finlay, 2002), the reciprocal approach to interdisciplinary AI research was developed based on research describing the history of convergence in GLAM, anecdotal conversations about interdisciplinary research within our own disciplines and inspired by our experiences as a social sciences academic and a computer science academic who have collaborated over several years on AI research funding proposals and in the joint-supervision of PhD students. Our aim is to provide recommendations that can be operationalised in collaborations but can also be introduced to early career researchers as part of formal and informal education and mentoring. This work is in progress, and we plan to further refine the framework through our reflexive approach. Future research will concretise the central characteristics of our reciprocal approach and provide example use cases.

## 4.1 Practice being open and upfront about jargon and disciplinary terms

To move past the jargon that can impede satisfactory interdisciplinary collaboration, we recommend that all researchers, regardless of discipline, reflect on how terms commonly used in their own disciplines may hold different meanings for researchers outside the respective discipline. This includes acknowledgement that disciplines contextualise these terms and there is no universally correct definition. Occasionally, definitions of common terms may be widely accepted within a specific discipline but unknown to those outside the domain. Some terms are ill defined yet commonly used. For example, consider the term “tool”, which may be utilised in the process of developing research objectives for an interdisciplinary collaborative project. In social science, tools may refer to a method, instrument, approach or framework. In contrast, within computing, a tool is more likely to refer to a digital application used to build or train models or to interact with systems.

In interdisciplinary research collaborations, it is important to address the use of terms and jargon upfront. We suggest that all members of the collaboration establish the significance of asking for clarity about terms and in exchange, be open to discussing what their use of different terms mean. This practice of reciprocal understanding is crucial for developing innovative approaches to implementing AHSS concepts and theories into ML and CV driven AI practice, especially if tension arises between different disciplinary norms. Affording time to consider these tensions and working through them can lead to jointly created best practices for development.

We consider this suggestion to be a reflexive practice, because it requires constant self-reflection on how a researcher's position within a discipline can impact the language that they use. This practice is relevant for the development and application of AI in different venues because it attunes the interdisciplinary team to the different terms relevant for AI use in different contexts. In addition, it may play a role in how early career researchers experience tension when faced with different terms and their willingness to interrogate use of those terms.

## 4.2 Interrogate how output, impact and novelty are defined and structured in ones' existing disciplines and how they shape interpretations of solutions

In our experience and in discussions within our own disciplines, individuals often seek out research collaborators when they acknowledge that they need additional expertise to address a research problem constructed through their own disciplinary lens. In this instance, we believe that the problem does not lie in the motivation to collaborate based on disciplinary norms, but lack of acknowledgement of the role that disciplinary norms play within the collaboration and lack of willingness to see beyond those norms. This disciplinary lens can shape interpretations of novelty, impact and indicators of achievement, such as targeted publication venues, journal reputation and methodological innovation. In the currently overcrowded research space on the topic of AI, what one discipline may find novel, another may find pedestrian or iterative. As a context, different approaches to AI research can also lead to the tension that Klumbyte et al. (2022) describes: conflicts between a slower pace preferred to address social implications and a faster pace employed to heighten pragmatic applicability.

We believe that reflexive practice can be applied to address this pace-related tension that different collaborators may experience, by affording space for researchers to notice the tension, reflect on it, and explore how it impacts the research process. For example, if a CV researcher were to read a humanities-based theory or practice that conflicts with disciplinary expectations of model performance, this may lead to tension if it conflicts with their previous understanding of evaluation metrics, particularly if the researcher was early in their career and focussed narrowly on "beating the start of the art performance". Within reflexive practice, the goal is not to eliminate tension, but rather to use it as a device for self-reflection on what we consider valid and valuable within research (Corlett & Mavin, 2018).

Reflecting on this tension mirrors the practice of GLAM professionals who must determine the need to maintain material that conflicts with their worldview. In a classic example from archival studies, Boles (1994) reflects on his experience as Director of a University aligned historical society that acquired Ku Klux Klan membership records into their archives. Using autoethnographic methods, Boles reflects on the internal conflict he experienced playing a role in the decision, which led him to engage with other GLAM professionals, and African American student organisations. Through these meetings, he described how he learned to listen to students while quieting his impulse to teach them about the process of acquisitions in archives. He concludes that the experience allowed him to develop new insight on how archivists jump to justify controversial acquisitions, when understanding the role multiculturalism plays in acquisition is often overlooked.

## 4.3 Develop a common understanding of the most pressing research problems in relation to AI

In describing GLAM convergence, Marty (2014) explains that for most people accessing cultural heritage material, they care little about whether a particular institution follows the practices aligned with library, archives or museum theory and/or concepts; they just want access to the information. The reality that the boundaries between GLAMs are often blurred for people trying to access materials supports the argument for inter and intra institutional convergence because the unifying concern of GLAMs should be to support users, not their own interests in adherence to specific work practices. For Marty, what unifies GLAMs is digital access; he urges professionals not to become preoccupied with drawing boundaries that are irrelevant for those outside the profession.

Disciplines trying to carve out their unique contributions in interdisciplinary collaborations can become overly concerned with reinforcing boundaries disregarded by those outside the specific discipline. A focus on expertise and knowledge remains relevant, but when focused on at an extreme, it can disrupt the ability of researchers to integrate their ideas, methods, approaches and worldviews in interdisciplinary research collaborations (Vienni-Baptista et al., 2024).

In AI research, integration is essential to tease out how perspectives define the research problem even in overlapping research areas. For example, the AHSS AI literature approaches AI as a societal problem in its discussion of ethics, fairness, bias and accountability. In contrast, computer vision and machine learning authored research is more likely to approach issues of ethics, fairness, bias and accountability as a problem with the behaviour, prediction and performance of computer programs (Eden, 2007). In this example, the problem of AI is classified as ethics, fairness, bias and accountability but the approach to the problem differs and could shape proposed solutions and accompanying research processes. Reflexive practice involves constant questioning of one's understanding of reality, including problems. To create novel approaches and joint research objectives, collaborators should make their understandings of the research problem explicit over several rounds of discussion, be open to questioning one's initial view of the problem and constantly scan for overlaps in each collaborator's views on the problem.

## 4.4 Identify values in relation to the use of AI in different contexts to jointly develop new approaches

Both Bastian (2017) and Marty (2014) describe how GLAM convergence was shaped by the adoption of novel technology to afford digital access to cultural heritage materials. Professional ethics, standards and best practices communicate a shared value of equitable public access to cultural heritage materials. This in turn shapes practices including acquisition, organisation, collection and preservation, that is extended to decisions to adopt technology that supports these practices. When observing decisions about whether or not to adopt AI systems to support cultural heritage work, Cushing and Osti (2023) found evidence that digital archivists have relied on these shared values, standards and best practices in determining if, how and when to utilise AI technology in their work.

This consideration for value differs significantly from technology acceptance models (TAMs) that often guide technology development. TAMs presume that when deciding whether or not to adopt a technology, users' considerations are largely guided by perceived ease of use and perceived usefulness (Marangunić & Granić, 2015). Dedication to a specific mission of equitable access and usefulness are both values worthy of inquiry and can lead to new insight about technology adoption when investigated *simultaneously.* In addition, evaluation methodologies for AI technologies generally rely on simple metrics such as accuracy and often fail to consider more holistic measures (Salaudeen et al., 2025; Ni et al., 2025). The value or impact of any resulting outputs ("tools") is therefore measured by this performance with a narrow goal of improving the "state of the art" benchmark. An interdisciplinary collaboration has the potential to improve our shared understanding of "usefulness", but it is not possible without acknowledging the different values relevant to the work, at the forefront of the project.

In relation to the development and application of AI research, a holistic approach to identifying these values is necessary. In its centring of individual experiences that shape the outcome of technology adoption, a reflexive approach prompts the researchers to discuss their different perceptions of the value of using technology in relation to professional mission(s), identifying overlaps and then gathering data about users lived experiences in the context of technology use. This process can identify and/or develop a new lens with which to view development and application of AI technology.

## 4.5 Consider the role of individual habits and beliefs in research collaborations

Vienni-Baptista et al. (2024) states that interdisciplinary research collaboration should aim to be sustainable. The role of individual habits, beliefs and personality should not be overlooked in the role they play in establishing sustainable interdisciplinary collaborations. While disciplinary norms can shape the worldview of researchers and their actions within interdisciplinary collaborations, it is important to remember that researchers from the same discipline are individuals with personalities that shape their working processes. The process of convergence in GLAMs has afforded space for these differences across institutions and disciplines by providing space for professionals and academics to interact and collaborate in disciplinary and joint academic and professional gatherings and projects (Marty, 2014).
Reflexivity affords critical self-reflection which can lead to a greater ability to know one's true self (Pillow, 2003). Knowing one's true self can afford identification of one's own habits and beliefs relevant in different contents. Within interdisciplinary collaborations, making these habits and beliefs explicitly early can allow researchers to identify collaborations that may be more sustainable over the long term.

# 5 Conclusion: A call to arms

Both society and research domains hold fractured and polarised views about the utility and impact of AI technologies. Therefore, greater understanding, empathy and exchange of views is critical to the success of collaborative research. We identify five key practices to enable reciprocal collaboration between AHSS and AI technology researchers using a guiding lens from the history of convergence in GLAM studies. Positioned as a framework to assist in the project development and execution phases, these practices can enhance communication and understanding between disciplines and identify substantial impacts and improvements. These practices can also guide the development of training for early career researchers. Current frameworks that guide doctoral education in the EU such as the Salzburg II Principles (European University Association, 2010), list the development of transferable skills and interdisciplinary training as central objectives in PhD Programmes. Practically, this involves translating disciplinary jargon, making different approaches explicit and acknowledging different disciplinary understandings of outputs and impacts. Future work is planned that will utilise reflexive and autoethnographic practises in PhD training to operationalise these concepts and refine their use in cutting edge research.